# Insights into Hydration Dynamics and Cooperative Interactions in Glycerol-Water Mixtures by Terahertz Dielectric Spectroscopy

Ali Charkhesht[1], Djamila Lou[1], Ben Sindle[1], Chengyuan Wen[1,2], Shengfeng Cheng[1,2], Nguyen Q. Vinh*,[1]

[1]Department of Physics and Center for Soft Matter and Biological Physics and [2]Macromolecules Innovation Institute, Virginia Tech, Blacksburg, Virginia 24061

Corresponding Author: *Email: vinh@vt.edu; Phone: 540-231-3158

## ABSTRACT

We report relaxation dynamics of glycerol-water mixtures as probed by megahertz-to-terahertz dielectric spectroscopy in a frequency range from 50 MHz to 0.5 THz at room temperature. The dielectric relaxation spectra reveal several polarization processes at the molecular level with different time constants and dielectric strengths, providing an understanding of the hydrogen-bonding network in glycerol-water mixtures. We have determined the structure of hydration shells around glycerol molecules and the dynamics of bound water as a function of glycerol concentration in solutions using the Debye relaxation model. The experimental results show the existence of a critical glycerol concentration of ~7.5 mol %, which is related to the number of water molecules in the hydration layer around a glycerol molecule. At higher glycerol concentrations, water molecules dispersed in a glycerol network become abundant and eventually dominate and four distinct relaxation processes emerge in the mixtures. The relaxation dynamics and hydration structure in glycerol-water mixtures are further probed with molecular dynamics simulations, which confirm the physical picture revealed by the dielectric spectroscopy.

## 1. INTRODUCTION

Along with water, a variety of co-solvents play important roles in biological systems.[1-5] The presence of co-solvents changes the behavior of water such as hydrogen-bonding network, dynamics, polar property, and spatial distribution.[6] Co-solvents can stabilize the activity of an enzyme and the native structure of a protein,[3] increase the aqueous solubility of a nonpolar drug by several orders of magnitude,[4] and enhance the chemical stability of a substance.[7] The study of the dynamics of these chemical biomolecules is indispensable to get a comprehensive perception of their conduct in aqueous solutions. Glycerol ($C_3H_8O_3$) is an important co-solvent in this context, which was a subject of numerous studies in molecular dynamics (MD) simulations[8-13] and experiments.[14-16] At ambient conditions, this trihydric alcohol with three hydroxyl groups is a colorless, sugar-like, highly viscous liquid. The high flexibility and viscosity of glycerol makes it an important system in studies of the glass transition.[17] Also, glycerol has been used to preserve proteins because of its cryoprotective properties,[18] and to stabilize enzyme activities.[2-3] The ability to form hydrogen bonds with water makes glycerol-water mixtures fascinating solutions for enhancing the solubility of several drugs.[4, 19] Thus, a comprehensive understanding of the hydration dynamics and cooperative interactions in glycerol-water mixtures is needed to help us understand the role of glycerol in these activities.

The investigation of molecular dynamics in a complex liquid is a major challenge in physical chemistry and chemical physics. Glycerol and glycerol-water mixtures have been the subject of numerous investigations including MD simulations,[10-12] broadband dielectric spectroscopy,[14, 17, 20-22] nuclear magnetic resonance (NMR),[23] infrared spectroscopy,[10] and Raman spectroscopy.[24] Although such a wide range of techniques has been employed to investigate the hydrogen-bonding dynamics of glycerol-water mixtures in different frequency ranges, the hydration dynamics in aqueous glycerol solutions is yet to be elucidated. The dielectric relaxation spectroscopy, which measures the rearrangement dynamics in a hydrogen-bond network, is a handy tool that can be used to advance our understanding of the hydration structure and dynamics in glycerol-water mixtures.

Recent developments in megahertz-to-terahertz spectroscopy provide us a possibility to conduct dielectric response measurements in a wide range of time scales to reveal hydration dynamics and cooperative interactions in glycerol-water mixtures. We have adopted this technique to investigate the structure and dynamics of hydration shells and the properties of water molecules interacting with proteins and micelles. The results enabled us to map out the physical behavior of different molecules in their aqueous solutions.[25-27] Our spectrometer covers a large spectral range from megahertz to terahertz frequencies and has a significantly improved signal-to-noise ratio with high power, providing high accuracy measurements.[28] In the present study, we focus on the nature of hydration dynamics and the molecular dynamics of glycerol in its aqueous solutions at the molecular level. From the complex dielectric response, we have explored the relaxation processes in these solutions that span a wide range of glycerol concentrations. The behavior of the hydrogen-bond network related to distinct relaxation times of bulk water, water molecules in a hydration layer around a glycerol, and water confined in a glycerol network has been discussed. A critical value of glycerol concentration is identified beyond which water molecules confined in a glycerol network start to emerge. The existence of different types of water with distinct relaxation times is further confirmed with molecular dynamics simulations. Finally, a unified physical picture is presented to help us understand the colligative properties of glycerol-water solutions.

## 2. EXPERIMENTAL METHODS

### 2.1. Materials.

Glycerol (≥ 99.5%) with molecular weight of 92.093 g/mol, purchased from Sigma Aldrich (Cat. No. 56-81-5), was used to prepare glycerol-water mixtures. The mixtures with glycerol content from 5 to 50% volume percentage with an increment of 5 vol % were prepared from the pure glycerol and deionized water (resistivity of 18.2 MΩ.cm). Measurements on pure glycerol and water were also performed, and the results are included in our discussion. Table 1 shows the glycerol volume percentage (vol %), weight-by-weight ratio (w/w), and a conversion to the glycerol molar percentage ($x_{glyc}$) of our glycerol-water mixtures.

### 2.2. Dielectric Spectroscopy.

Measuring dielectric relaxation properties of glycerol-water mixtures at megahertz-to-terahertz frequencies provides insights into the structure and dynamics of these dipolar liquids. Our spectrometer allows us to study the relaxational (rotational) as well as translational motion of water and glycerol molecules. The technique is absolutely essential to extract different dynamics of water molecules in bulk water, hydration layers, and a glycerol network, and to probe the relaxation process of glycerol.

The dielectric relaxation spectroscopy of a liquid is a powerful tool to reveal different dynamical processes at the molecular level. The dielectric spectroscopy of glycerol-water mixtures in a frequency range from 10 μHz to 30 GHz at temperature from 148 to 323 K was performed by Hayashi et al.[20, 29] and Puzenko et al.,[17] where both water-rich and glycerol-rich regions were probed[22]. The analyses were focused on the dielectric loss in both regions, and they used well-known phenomenological relations and their superposition for data fitting. They concluded that the main dielectric relaxation process, the high-frequency "excess wing", and the dc conductivity in glycerol-water mixtures have the same origin. A schematic model was provided for different relaxation processes of molecules in glycerol solutions, resulting from ice nanocrystals or pure water (w-w interactions), pure glycerol (g-g interactions), and glycerol-water complexes (g-w interactions). However, due to the limitation of the frequency range probed, these studies did not reveal exclusive details about the hydration layer dynamics that would be placed at higher frequencies, the "excess wing". Alternatively, Dashnau et al. performed infrared spectroscopy on glycerol-water mixtures to study hydrogen-bond patterns and cryoprotective properties at various concentrations.[10]

They discussed how the properties of the hydrogen-bond network and hydration shells change when the glycerol concentration is increased. The properties were determined using the Fourier-transform infrared spectroscopy (FTIR) and MD simulations that show that stretch modes of CH and OH bonds depending on glycerol concentration. Using our dielectric spectroscopy covering megahertz-to-terahertz frequencies, we aim to reveal a physical picture of the relaxation dynamics of water molecules in a hydration shell enclosing a glycerol molecule as well as those strongly confined between glycerol molecules, i.e., dispersed in a glycerol network.

At microscopic scales, numerous polarization effects give rise to the dielectric properties of a mixture. Glycerol/biomolecules and water molecules with permanent dipole moments rotate to follow an alternating electrical field from a radiation source. Each dielectric mechanism has a characteristic frequency. In the megahertz-to-terahertz frequency range, electronic and atomic polarization mechanisms are comparatively weak, and make a constant contribution to the measured signal. In this range of frequencies, the dielectric response of an aqueous solution is mainly controlled by three processes: (*a*) the rotational motion of solutes such as glycerol and biomolecules, i.e., the orientational polarization of the solute dipoles; (*b*) the orientational relaxation of bulk water molecules, i.e., water dipoles; and (*c*) the relaxation process of water molecules in a hydration shell surrounding a glycerol molecule or a biomolecule, i.e., the relaxation of the dipoles of water molecules in a hydration layer, or confined water molecules in a glycerol network.[25-26, 28, 30]

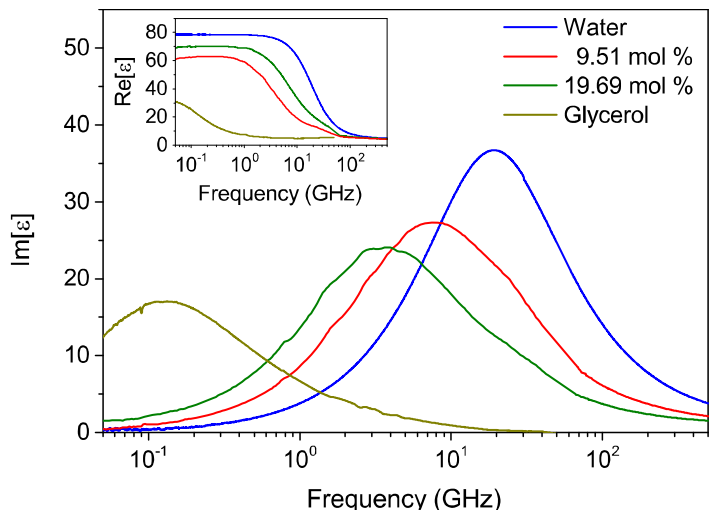

**Figure 1:** Interaction of electromagnetic wave in the megahertz-to-terahertz region with glycerol-water mixtures providing insight into the molecular dynamics over the picosecond to sub-microsecond timescales. The imaginary, $\epsilon''_{sol}(\nu)$, and the real, $\epsilon'_{sol}(\nu)$, (in the inset) components of the dielectric response spectra were collected for aqueous glycerol solutions at various glycerol concentrations. The maximum of the imaginary component centered at ~ 19.2 GHz for pure water moves to lower frequencies for glycerol-water mixtures, and stays at ~ 144.7 MHz for pure glycerol liquid.

To examine the relaxation processes of water and glycerol in solutions, measurements of the complex dielectric function for each sample were carried out in a large frequency range from 50 MHz to 0.5 THz at 25 ºC. In the frequency range from 50 MHz to 50 GHz, we use an enhanced open-end probe (Agilent 85070E) and a vector network analyzer (Agilent PNA N5225A). The calibration of this system was performed under three standards including air, pure water, and mercury (short circuit). The complex dielectric response including the real (dielectric dispersion), $\epsilon'_{sol}(\nu)$, and the imaginary (dielectric loss), $\epsilon''_{sol}(\nu)$, components was evaluated using Agilent software with an accuracy of $\Delta\epsilon/\epsilon = 0.05$. In the frequency range from 60 GHz to 0.5 THz, the dielectric response of glycerol-water mixtures has been collected using a gigahertz-to-terahertz spectrometer based on the above vector network analyzer with frequency extenders from Virginia Diodes. Our spectrometer is capable of simultaneously measuring intensity and phase over a large effective dynamical range.[28] The solutions were kept in a sample cell made of anodized aluminum at 25 °C and the temperature was controlled with an accuracy of ± 0.02 °C using a Lakeshore 336 temperature controller.

The complex dielectric response of a glycerol mixture can be expressed as a function of frequency, ν:

$$\epsilon^*_{sol}(\nu) = \epsilon'_{sol}(\nu) + i\epsilon''_{sol}(\nu) \qquad (1)$$

Figure 1 shows the complex dielectric response spectra for the mixtures with different glycerol concentrations from 0 mol % (pure water) to ~ 20 mol %. The dielectric spectra of pure glycerol and pure water are also included as reference. When the concentration of glycerol is increased, the absorption of the mixture decreases dramatically and the maximum of the dielectric loss shifts significantly toward lower frequencies. These changes are expected to occur as a water molecular and a glycerol molecule have different molecular weight and polarity, which affect the orientational relaxation of the associated dipoles.[17, 20, 25-26]

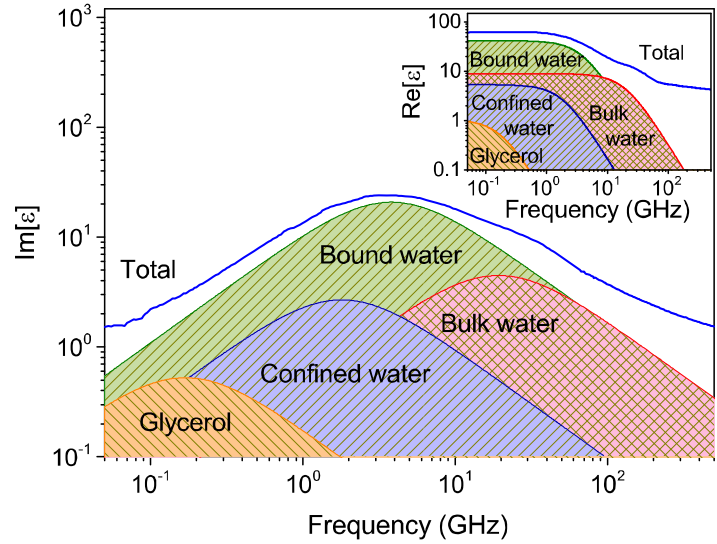

**Figure 2:** Dielectric response of the 19.69 mol % glycerol-water mixture in the frequency range from 50 MHz to 0.5 THz reflecting the complexity of glycerol-water interactions. The imaginary and the real (in the inset) components of the dielectric spectra have been decomposed in to four relaxational processes with different relaxation time constants.

### 3. RESULTS AND DISCUSSION

The dielectric properties of an aqueous solution present a complex behavior, originating from different, and/or partially overlapping, polarization mechanisms. In order to determine the contribution of different components in a solution to its dielectric response, the data were analyzed by simultaneously fitting the measured real, $\epsilon'_{sol}(\nu)$, and imaginary, $\epsilon''_{sol}(\nu)$, components to the Debye relaxation model based on a sum of four individual contributions:[31]

$$\epsilon^*_{sol}(\nu) = \epsilon_\infty + \frac{\epsilon_S - \epsilon_1}{1 + i2\pi\nu\tau_1} + \frac{\epsilon_1 - \epsilon_2}{1 + i2\pi\nu\tau_2} + \frac{\epsilon_2 - \epsilon_3}{1 + i2\pi\nu\tau_3} + \frac{\epsilon_3 - \epsilon_\infty}{1 + i2\pi\nu\tau_4} \qquad (2)$$

where $\Delta\epsilon_1 = \epsilon_S - \epsilon_1$, $\Delta\epsilon_2 = \epsilon_1 - \epsilon_2$, $\Delta\epsilon_3 = \epsilon_2 - \epsilon_3$ and $\Delta\epsilon_4 = \epsilon_3 - \epsilon_\infty$ are the associated dielectric strengths, corresponding to four relaxation times $\tau_1, \tau_2, \tau_3$, and $\tau_4$. Each relaxation time corresponds to a characteristic relaxation frequency via $\tau_j = 1/(2\pi\nu_j)$. The static permittivity, $\epsilon_s$, is given by $\epsilon_s = \epsilon_\infty + \sum_{j=1}^{4}\Delta\epsilon_j$. The value of $\epsilon_\infty$ indicates the dielectric contribution from all modes at frequencies much higher than the range probed in our dielectric measurements, such as molecular and atomic oscillations. In the megahertz-to-terahertz frequency range, librational motions and inertial effects all together made a constant contribution to the dielectric response, captured by $\epsilon_\infty$.

Both dielectric dispersions, $\epsilon'_{sol}(\nu)$, and loss, $\epsilon''_{sol}(\nu)$, are fitted to Eq. (2) concurrently with a set of free parameters. The relaxation time for bulk water, $\tau_4 \sim 8.27$ ps,[32] ($\nu_4 \sim 19.2$ GHz) and pure glycerol, $\tau_1 \sim 1100$ ps[22] ($\nu_1 \sim 144.7$ MHz) are used as initial conditions. A deconvolution of the dielectric spectra including the real and imaginary components has been performed for the glycerol-water mixture with molar percentage, $x_{glyc}$, of 19.69 mol %. The complex dielectric spectra in Fig. 2 for this mixture indicate that the four relaxation processes are centered at $168 \pm 18$ MHz ($\tau_1 \approx 945 \pm 115$ ps), $1.8 \pm 0.2$ GHz ($\tau_2 \approx 88 \pm 9$ ps), $4.0 \pm 0.7$ GHz ($\tau_3 \approx 39 \pm 8$ ps), and $19.2 \pm 0.8$ GHz ($\tau_4 \approx 8.27 \pm 0.35$ ps). Corresponding to the four components of the imaginary part, the deconvolution for the real part has been shown in the inset of Figure 2. The dielectric constant at higher frequencies obtained from this fitting procedure, $\varepsilon_\infty = 7.06 \pm 0.72$, is within experimental uncertainty of the value reported in the literature.[32-34] The fitting also yields the values of the dielectric strength, $\Delta\varepsilon_1 = 1.35 \pm 0.05$, $\Delta\varepsilon_2 = 1.26 \pm 0.27$, $\Delta\varepsilon_3 = 44.47 \pm 2.10$, and $\Delta\varepsilon_4 = 8.95 \pm 0.45$. The relaxation times and the dielectric strengths at different glycerol concentrations are reported in Figs. 3 to 5 and Table 1. The slowest relaxation time, $\tau_1 \approx 910 \pm 135$ ps, in average, originates from the dynamics of glycerol in aqueous solutions. The relaxation time of glycerol in aqueous solutions is only slightly faster than those in pure glycerol, indicating that the dynamics of glycerol in an aqueous solution remains close to those in a pure glycerol. The fastest relaxation time, $\tau_4 \approx 8.27$ ps, comes from bulk water in the solution. The relaxation times, $\tau_2$ and $\tau_3$, can be assigned to the water molecules confined in a glycerol network and those in the hydration layer around a glycerol molecule, respectively. This physical picture of dielectric relaxation in aqueous glycerol solutions is discussed in detail below and further confirmed with molecular dynamics simulations.

### 3.1. Relaxation of glycerol in aqueous solutions

The dielectric parameters of glycerol-water mixtures allow us to evaluate relaxational processes of molecules in the mixtures. Interestingly, the relaxation times for the four processes identified by fitting the dielectric relaxation spectra to the Debye model are almost constant when the glycerol concentration is varied. The slowest relaxation time, $\tau_1$, is around $910 \pm 135$ ps or $174 \pm 22$ MHz (Figure 3, inset). This mode is called the $\beta$-relaxation, which is a typical relaxation process in glass-forming liquids. This value is slightly faster than those in pure glycerol.[17, 35-36] Furthermore, the dielectric strength of this relaxation process increases approximately linearly with the glycerol concentration (Figure 3, inset). The indication is that glycerol molecules form strong hydrogen bonds with water molecules in an aqueous solution and these bonds are similar to the hydrogen bonds between glycerol molecules in a pure glycerol. As a result, glycerol in an aqueous solution exists in an environment which is not very different from that in a pure glycerol as far as its rotational motion is concerned.

The dielectric parameters of the $\beta$-relaxation process, including the dielectric strength, $\Delta\varepsilon_1$, and the relaxation time, $\tau_1$, allow us to evaluate the hydrodynamic radius of a glycerol molecule and its electric dipole moment, $\mu$, in glycerol-water mixtures. In a simple physical model, a dipole is regarded as a sphere whose rotation in response to an electric field is opposed by the hydrodynamic friction with the surrounding solvent. The relaxation time of a spherical molecule in a diluted polar solution with hydrodynamic radius, $R$, rotating in a medium with macroscopic viscosity, $\theta$, is given by the Debye equation[37]

$$\tau_1 = \frac{4\pi R^3 \theta}{k_B T} \tag{3}$$

where $k_B$ is the Boltzmann constant, $T$ is temperature. At 25 °C, the viscosity of water is about $0.89 \times 10^{-3}$ kg·m$^{-1}$·s$^{-1}$ (Pa·s).[38] From the measured value of $\tau_1$, we derive the hydrodynamic radius of glycerol in low glycerol concentration solutions as $R = 7.06$ Å. Note that this radius is larger than the unit cell dimensions of a glycerol molecule ($a = 4.8$ Å, $b = 5.1$ Å, $c = 7.8$ Å)[38-39]. This difference is expected as the hydrodynamic radius estimated from the dielectric measurements includes the contribution from the bound water molecules around a glycerol molecule.

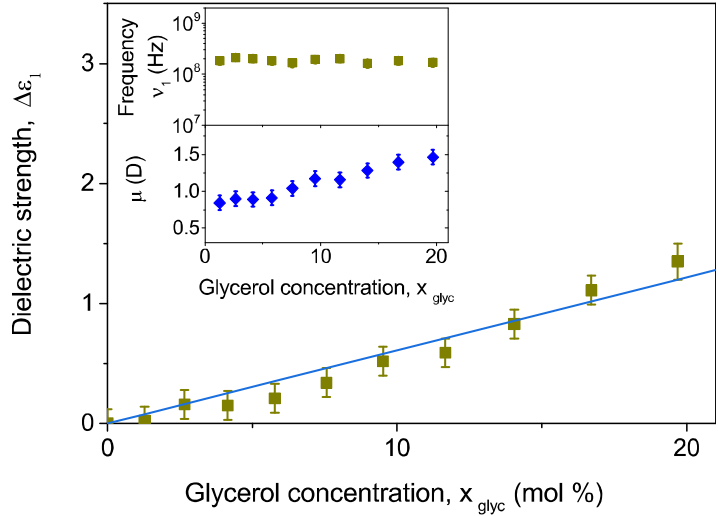

**Figure 3:** Results of dielectric relaxation response revealing the existence of several relaxation modes in glycerol-water mixtures. While the relaxation frequency (upper inset) of glycerol, $\nu_1$, is almost constant as the glycerol concentration is varied, the dielectric strength, $\Delta\epsilon_1$ ($\nu$), of glycerol-glycerol interaction increases with an increasing glycerol concentration. In the lower inset, the values of the effective dipole moment of a glycerol molecule in the mixtures have been estimated from the dielectric response.

The effective dipole moment of glycerol in an aqueous solution can be estimated from the dielectric strength. Several approximations for such estimation have been suggested. We adopted the same approach as in Ref. [40] to calculate the effective dipole moment of glycerol, $\mu_{\text{eff}}$, by using the Onsager–Oncley model[41]

$$\mu_{\text{eff}}^2 = \frac{2\varepsilon_0 k_B T \Delta\varepsilon_1}{N_A c g_K} \qquad (4)$$

where $N_A$ is the Avogadro's number, $c$ is the molar concentration (mol/m$^3$) of glycerol in the solution, $\varepsilon_0$ is the permittivity of vacuum, and $g_K$ denotes the Kirkwood correlation factor and is often assumed to be 1 in a diluted solution.[37, 42] The estimated values of the effective dipole moment of glycerol at various concentrations are shown in the lower inset of Figure 3. At a low glycerol concentration, $\mu_{\text{eff}}$ is roughly a constant of 0.85 D, and it starts to increase with an increasing glycerol concentration when the concentration is higher than about 7.5 mol %. At low concentration, each glycerol molecule is well covered by a hydration layer that is separated from other hydration layers. When the glycerol concentration increases beyond the threshold, glycerol molecules tend to form clusters or networks and hydration layers start to overlap. Since the dipole moment of glycerol is 2.56 D, an increase of the effective dipole moment of glycerol is observed as the glycerol concentration is increased further.

### 3.2. Bulk water relaxation and hydration effect

The dielectric spectroscopy provides insights into the dynamics of bulk water, water bound to glycerol, and water confined in a glycerol network. The relaxation time for bulk water, $\tau_4 \approx 8.27$ ps, is independent of glycerol concentration and similar to the values reported in the literature for pure water at gigahertz frequencies.[32-33] However, when glycerol is added to a glycerol-water mixture, the dielectric strength, $\Delta\varepsilon_4$, of the bulk water in the mixture (Figure 4a) decreases significantly. The lowering of the dielectric response with an increasing glycerol concentration comes from two main reasons. Firstly, the presence of glycerol will reduce the concentration of water in the mixture, thus lowering the dielectric response of water. Secondly, water molecules form hydrogen-bonds with glycerol. A glycerol molecule has three OH groups and can form 6 hydrogen bonds with surrounding water in its hydration layer. When the glycerol concentration is increased, the amount of water in the hydration layers also increases. These water molecules have a different relaxation process than the bulk water. As a result, the dielectric strength of the relaxation mode of bulk water in the mixture decreases as the glycerol concentration is increased.

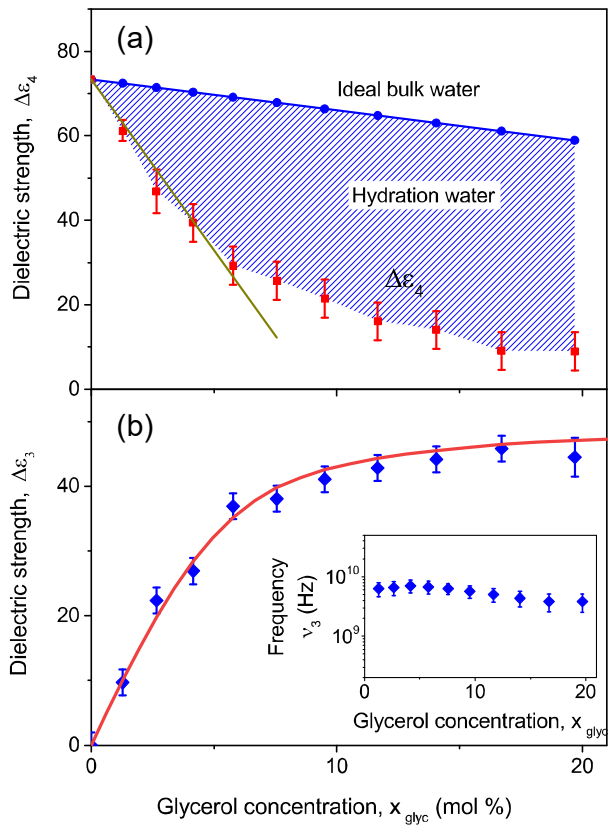

**Figure 4:** Dielectric spectra of glycerol mixtures revealing the number of water molecules affected by the presence of glycerol. (a) The dielectric strength of bulk water in glycerol, $\Delta\epsilon_4$ ($\nu$), decreases significantly with an increasing glycerol concentration. The solid line (blue) represents the dielectric strength of the "*ideal bulk water*" extracted by assuming that all water molecules in the mixtures behave as pure water and relax with the time constant of 8.27 ps. The straight line in the low concentration region is a guide for the eye. (b) Amplitude of the dielectric property of the bound water in glycerol-water mixtures increases with an increasing glycerol concentration. The solid line in red color is a guide for the eye. In the lower inset, the relaxation frequency of bound water in the hydration layer is almost constant as the glycerol concentration is varied.

The average number of water molecules in the hydration layer of a glycerol molecule as a function of glycerol concentration is an important quantity in the understanding of colligative properties of glycerol-water mixtures. To reveal the hydration structure, we compute the dielectric strength of a mixture by assuming that all water molecules in the mixture are "*ideal bulk water*" and take part in the pure water relaxational mode. Figure 4a shows the contribution (the solid blue line) of such "*ideal bulk water*" to the dielectric response of the glycerol-water mixtures as a function of glycerol concentration. However, the measured dielectric response of water in the mixtures, $\Delta\varepsilon_4$, from the bulk relaxation mode is lower than this estimate. That is, the "*experimental bulk water*" is less than the "*ideal bulk water*", and not all water molecules in a mixture participate in the bulk water relaxational process characterized by the relaxation time, $\tau_4$. The difference between the estimated dielectric response and the measured one can be used to extract the number of water molecules missing from the pool of bulk water. These water molecules form hydrogen-bonds with glycerol molecules and relax with different characteristics.

**Table 1:** Glycerol-water mixtures used for the MHz-to-THz dielectric spectroscopy measurements, and dielectric relaxation times, $\tau_1$, $\tau_2$, and $\tau_3$, for these mixtures at various concentrations at 25 °C in which the relaxation of bulk water, $\tau_4$, is $8.27 \pm 0.35$ ps.

| Glycerol volume percentage (vol %) | Weight ratio, (w/w) | Glycerol molar percentage, $x_{glyc}$ (mol %) | $\tau_1$ (ps) | $\tau_2$ (ps) | $\tau_3$ (ps) | $\Delta\varepsilon_1$ | $\Delta\varepsilon_2$ | $\Delta\varepsilon_3$ | $\Delta\varepsilon_4$ |
|---|---|---|---|---|---|---|---|---|---|
| 0 | 0.000 | 0 | -- | -- | -- | -- | -- | -- | 73 |
| 5 | 0.066 | 1.27 | 815 | 77 | 27 | 0.02 | 0.05 | 9.69 | 61.22 |
| 10 | 0.139 | 2.65 | 875 | 75 | 26 | 0.16 | 0.10 | 22.34 | 45.87 |
| 15 | 0.221 | 4.15 | 890 | 79 | 29 | 0.15 | 0.12 | 26.89 | 39.44 |
| 20 | 0.313 | 5.78 | 975 | 77 | 28 | 0.21 | 0.15 | 36.90 | 29.22 |
| 25 | 0.418 | 7.56 | 955 | 86 | 34 | 0.34 | 0.21 | 38.04 | 25.69 |
| 30 | 0.537 | 9.51 | 880 | 78 | 36 | 0.52 | 0.61 | 41.12 | 21.44 |
| 35 | 0.675 | 11.66 | 890 | 81 | 32 | 0.59 | 0.76 | 42.85 | 16.09 |
| 40 | 0.836 | 14.05 | 972 | 83 | 36 | 0.83 | 0.75 | 44.18 | 14.02 |
| 45 | 1.026 | 16.71 | 870 | 87 | 38 | 1.11 | 1.34 | 45.83 | 9.03 |
| 50 | 1.254 | 19.69 | 945 | 88 | 39 | 1.35 | 1.26 | 44.47 | 8.95 |
| 100 | ∞ | 100 | 1100 | -- | -- | 37 | -- | -- | -- |

The relaxation dynamics of water molecules around a glycerol can be extracted from our analysis on the basis of the Debye model. Specifically, two relaxation processes with a frequency of $4.5 \pm 0.9$ GHz (Figure 4b, inset) and another one at $1.87 \pm 0.22$ GHz (Figure 5, inset), corresponding to the time constants of $35 \pm 8$ ps and $85 \pm 9$ ps, respectively, are identified. These time constants are much longer than that of pure water, which is 8.27 ps. As confirmed later with molecular dynamics simulations, the relaxation time, $\tau_3 \approx 35 \pm 8$ ps, can be associated with water molecules bound to a glycerol molecule and forming its hydration layer, whereas the even longer relaxation time, $\tau_2 \approx 85 \pm 9$ ps, can be assigned to water molecules confined in a glycerol network, i.e., water molecules shared by more than one glycerol molecules. The longer relaxation times of the bound and confined water indicate that the hydrogen-bond between a glycerol molecule and a water molecule is stronger than that between two water molecules.

The amount of bound water molecules that do not take part in the bulk water rotational process in a mixture and instead relax with a longer relaxation time, $\tau_3$, can be estimated from the dielectric strength, $\Delta\epsilon_3$, of the corresponding relaxation process. The hydration number, which denotes the average number of water molecules in the hydration layer of a glycerol molecule, is given by [26, 43-45]

$$N_{\text{hyd}}(c) = \frac{c_w - \frac{\Delta\epsilon_3}{\Delta\epsilon_{\text{pure}}} c_{\text{pure}}}{c} \qquad (5)$$

where $c$ is the glycerol molar concentration, $c_w$ is the molar concentration of water in the mixture, and $c_{\text{pure}}$ = 55.35 M is the molarity and $\Delta\epsilon_{\text{pure}} = 73.25$ is the dielectric strength of pure water at 25 °C.[32, 46]

The dielectric strength of the bound water molecules varies non-linearly with glycerol concentration, $x_{\text{glyc}}$. At low glycerol concentrations ($0 < x_{\text{glyc}} < 10$ mol %), the dielectric strength of the bound water increases linearly with glycerol concentration. When the glycerol concentration is higher than about 10 mol %, the dielectric strength shows a saturation behavior. In the low-concentration regime, our analysis shows that by average 5.58 water molecules are present in the hydration shell around a glycerol molecule, which agrees well with the average number of water molecules in the primary hydration layer of glycerol, 5.57, directly computed in MD simulations for low-glycerol-concentration solutions. This number is expected as each glycerol molecule can form at most 6 hydrogen bonds with surrounding water molecules through its 3 OH groups. The value also agrees with other reports in the literature.[10, 47] In a dilute mixture, glycerol molecules are uniformly dispersed in the mixture. The average number of water molecules bound to a glycerol molecule is roughly constant as long as the hydration shells of different glycerol molecules do not overlap. When the glycerol concentration is increased beyond a certain value, the hydration layers start to overlap and glycerol molecules aggregate to form clusters and networks, resulting in a decrease in the hydration number. Since the dielectric response of the bound water shows a saturation behavior at the glycerol concentration of ~7.5 mol %, this concentration is roughly the threshold value signaling the overlapping of hydration shells. Observations of a similar transition of the hydration structure have been reported for aqueous solutions of bovine serum albumin,[25] lysozyme proteins,[27, 43] and micelles.[26]

### 3.3. Confined water in the glycerol network

Water and glycerol molecules are well associated in mixtures with high glycerol concentrations. Dielectric spectra of high glycerol concentration solutions suggested that water cooperative domains do not exist and water molecules are dispersed well in the mixtures.[17, 22] The long relaxation time extracted from our dielectric response measurements (Figure 2), $\tau_2 = 85 \pm 9$ ps, corresponding to a characteristic frequency of $1.87 \pm 0.22$ GHz, is for water confined in a glycerol network and strongly bound to more than one glycerol molecules. In other words, these water molecules are in overlapped hydration shells. Since these water molecules have strong hydrogen-bonds with more than one glycerol molecules, they relax much more slowly. Furthermore, this relaxation time is almost constant as the glycerol concentration is varied (Figure 5, inset). In a mixture with a low glycerol concentration, the contribution of the confined water to the overall dielectric response of the mixture is negligible, indicating that the amount of confined water is insignificant. When the glycerol concentration is higher than about 7.5 mol %, hydration shells of different glycerol molecules start to overlap and the confined water emerges. This value coincides with that observed for the saturation value of the dielectric strength of bound water in Figure 4b, and the increasing of the effective dipole moment of glycerol in solutions (Figure 3, lower inset). After that point, the amount of confined water grows almost linearly with the glycerol concentration, as indicated by the dielectric strength, $\Delta\epsilon_2$, of the corresponding relaxation process (Figure 5). In a glycerol-rich mixture, this category of water dominates and the dielectric response of the mixture is mainly controlled by the glycerol network and the water confined in it. This physical picture motivates us to perform molecular dynamics simulations of glycerol-water mixtures and extract relaxation times of water as discussed below.

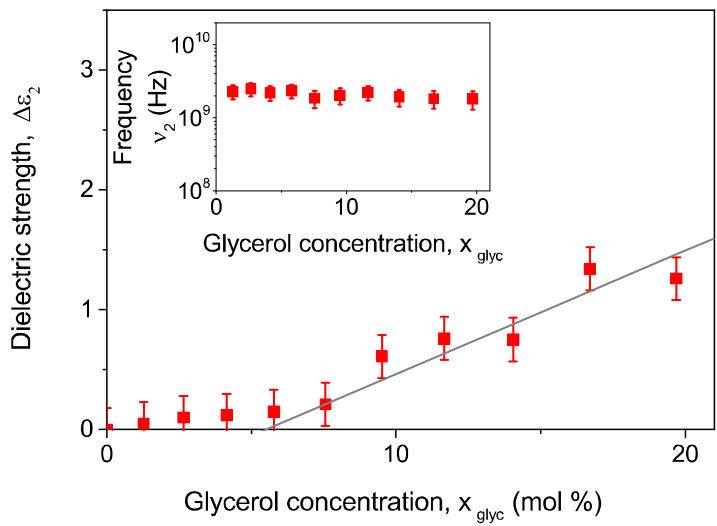

**Figure 5:** A slow dynamics of confined water molecules in the glycerol network emerging in the dielectric response of glycerol-water mixtures. Below 7.5 mol %, the contribution the confined water molecules to the dielectric response of glycerol-water mixtures is negligible. Beyond the critical concentration, the dielectric strength increases linearly with an increasing glycerol concentration. The solid line is a guide for the eye. In the inset, the relaxation frequency of confined water in the glycerol network is almost constant as the glycerol concentration is varied.

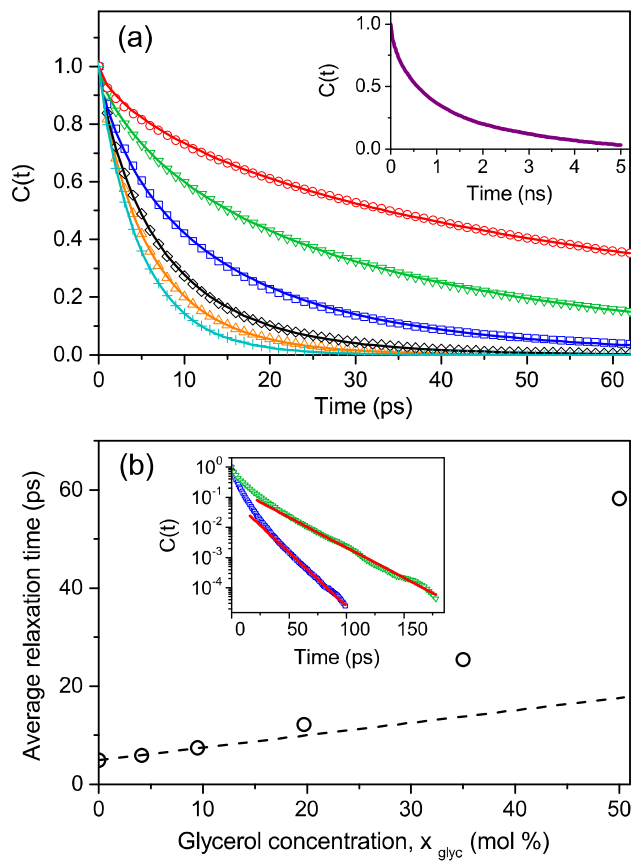

**Figure 6:** Relaxation autocorrelation functions, $C(t)$, for water and glycerol molecules in glycerol-water mixtures showing multiple-exponential decay behavior. (a) The relaxation autocorrelation functions and their fitting curves to a stretched-exponential function for pure water and 5 glycerol-water mixtures with glycerol concentration of 0 (cyan pluses), 4.15 (orange upward triangles), 9.51 (black diamonds), 19.69 (blue squares), 35 (green downward triangles), and 50 mol % (red circles) provide insight into the dynamics of water in the solutions. The relaxation autocorrelation function of pure glycerol is shown in the upper inset. (b) The average relaxation time shows a deviation from a linear dependence (dash line) on the glycerol centration when it is higher than about 10 mol %. The relaxation autocorrelation functions are plotted on a log-linear scale (lower inset) for glycerol-water mixtures with glycerol concentration at 19.69 mol % (blue squares) and 35 mol % (green downward triangles); the bottom (top) solid line has a slope that corresponds to a relaxation time of about 28 ps (50 ps).

### 3.4. Molecular Dynamics Simulations

To further probe the dynamics of water and glycerol at the molecular scales, we conducted all-atom MD simulation of their mixtures. All simulations were performed with Large-scale Atomic/Molecular Massively Parallel Simulator (LAMMPS).[48] Totally 7 systems with glycerol molar fraction ranging from 0 (pure water) to 1 (pure glycerol) were simulated and their parameters are listed in Table 2. The structure of a glycerol molecule was constructed using Automated Topology Builder.[49] The GROMOS force field[50] was adopted for glycerol and the SPC/E model[51-52] was used for water. The cut off of the 12-6 Lennard-Jones potential was set as 14 Å for glycerol and 7.9 Å for water. Coulomb interactions were fully accounted for with the long-range part computed with the particle-particle particle-mesh (PPPM) method. Geometric mixing rule was adopted for glycerol-water interactions. The equations of motions were integrated with a velocity Verlet algorithm with a time-step of 1 fs. Each system was equilibrated in an NPT ensemble at 1 atmospheric pressure and 300 K for 10 ns. After the density of each system became stable, an NVT ensemble was used for production runs with the system temperature fixed at 300 K using a Nose-Hoover thermostat. For the pure glycerol system, the production run was 10 ns and a snapshot of the system was dumped every ps. For all other systems, the production run was 500 ps and a snapshot was dumped every 0.05 ps.

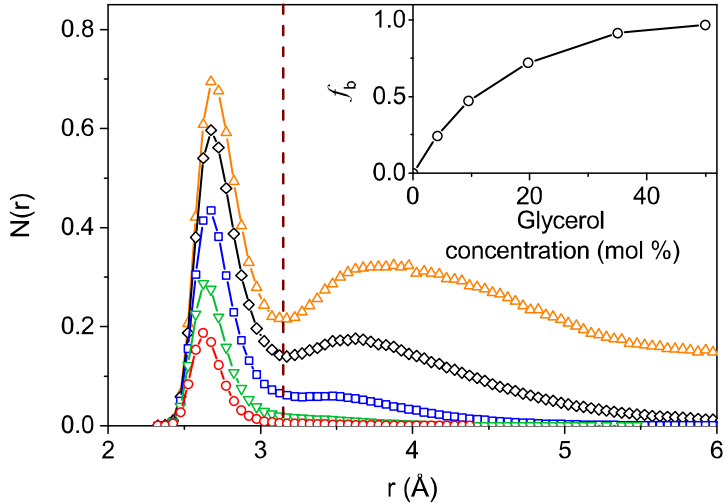

**Figure 7:** Normalized hydration number functions, $N(r)$, in the shell from $r - \delta r$ to $r + \delta r$ with $\delta r = 0.025$ Å, where $r$ is the shortest distance between the oxygen atom on a water molecule and an oxygen or a carbon atom on a glycerol molecule. Data are collected for glycerol concentration of 4.15 mol % (orange upward triangles), 9.51 mol % (black diamonds), 19.69 mol % (blue squares), 35 mol % (green downward triangles), and 50 mol % (red circles). The inset shows the fraction of water in the primary hydration shell of any glycerol molecules, $f_b$, as a function of glycerol concentration.

We calculated the relaxation autocorrelation function, $C(t)$, of a dipole using Eq. (6) in which $\mu_i(t)$ is the unit electric dipole associated with a water or a glycerol molecule at time $t$, and the summation is over all $N$ molecules (either water or glycerol) in a system. An ensemble average was computed with various states of the system being taken as the initial state at $t = 0$ to get the autocorrelation function

$$C(t) = \langle \frac{1}{N} \sum_{i=1}^{N} \mu_i(t)\, \mu_i(0) \rangle \qquad (6)$$

An analysis indicates that all relaxation autocorrelation functions can be fit to a stretched-exponential form, $e^{-(t/\tau)^\beta}$, with $\tau$ being an average relaxation time and $\beta$ being an exponent. When $\beta = 1$, the relaxation is exponential. The relaxation autocorrelation functions as well as their fits for pure water, 5 glycerol-water mixtures are shown in Figure 6a, and the results for pure glycerol is presented in the inset of Figure 6a. All results on $\tau$ and $\beta$ from such a fit are included in Table 2. For pure water, $\beta$ is close to 1 and the relaxation time is about 4.9 ps, which is very close to the value reported in the literature.[53] For pure glycerol, $\beta$, is about 0.69 and the relaxation time is about 986 ps. These relaxation time values are in good agreement with the dielectric spectroscopy measurements, ~8 ps for bulk water and ~1100 ps for glycerol.

The average relaxation time for water extracted from an autocorrelation function provides valuable information on the dynamics of water molecules in a glycerol mixture. As the glycerol concentration is increased, the average relaxation time of water increases (Figure 6b), indicating that water molecules in the hydration shell of a glycerol molecule have strong hydrogen bonds with the glycerol. As a result, the orientations of water molecules bound to glycerol fluctuate more slowly than those in bulk water. When the glycerol concentration is increased from 0 to 10 mol %, the average relaxation time has an approximately linear dependence on the glycerol concentration. This trend indicates that water in these systems can be effectively treated as a mixture of two types of water including bulk water and water bound to glycerol molecules. The latter has a longer relaxation time and its share increases with respect to bulk water as the glycerol concentration is increased. Specifically, the relaxation time for bound water can be extracted from an analysis of mixtures with high glycerol concentration of 19.69 and 35 mol %. The relaxation autocorrelation functions, $C(t)$, are plotted on a log-linear scale for this mixture (inset of Figure 6b). On short time scales, the relaxation is dominated by bulk water. However, on longer time scales the relaxation is almost exponential as indicated by the linear decay of $C(t)$ in the region of 30 ps $< t <$ 100 ps. The corresponding relaxation time is about 28 ps, which is only slightly lower than the experimental value of 35 ps identified from the dielectric spectroscopy data. The difference can be partially attributed to the SPC/E model adopted here for water. For bulk water, a similar trend can be noted that the relaxation time from the SPC/E model is 4.9 ps, which is also lower than the experimental value of ~8 ps.

**Table 2:** Glycerol-water mixtures studied in MD simulations. The first 4 columns show the molar fraction of glycerol, the number of water molecules, the number of glycerol molecules, and the size of the cubic simulation box for each system. $N_{\text{hyd}}$ is the number of bound water molecules per glycerol molecule, which are defined as the water molecules in the primary hydration shell of a glycerol molecule. $f_b$ is the fraction of water in each system as the bound water and mathematically, $f_b = N_{\text{hyd}} \times N_{\text{glyc}}/N_{\text{water}}$. The relaxation autocorrelation functions, $C(t)$, are fitted to a stretched-exponential form $e^{-(t/\tau)^\beta}$ with $\beta$ being an exponent and $\tau$ being an average relaxation time. Here we show the average relaxation time of water in each solution except in pure glycerol, where $\tau$ indicates the relaxation time of glycerol.

| Mol % | $N_{\text{water}}$ | $N_{\text{glyc}}$ | $L$ (Å) | $N_{\text{hyd}}$ | $f_b$ (%) | $\tau$ (ps) | $\beta$ |
|---|---|---|---|---|---|---|---|
| 0 | 2887 | 0 | 44.4 | 0 | 0 | 4.9 | 0.93 |
| 4.15 | 2887 | 125 | 46.7 | 5.57 | 24 | 5.9 | 0.88 |
| 9.51 | 2055 | 216 | 44.3 | 4.48 | 47 | 7.4 | 0.83 |
| 19.69 | 2088 | 512 | 49.6 | 2.94 | 72 | 12.2 | 0.75 |
| 35 | 951 | 512 | 44.2 | 1.70 | 92 | 25.4 | 0.72 |
| 50 | 512 | 512 | 41.9 | 0.97 | 97 | 58.2 | 0.67 |
| 100 | 0 | 1000 | 48.4 | -- | -- | 986 | 0.69 |

To better understand the structure of hydration shells around glycerol molecules, we analyze the distribution of water molecules when the glycerol concentration is varied. For this study, we consider the position of the oxygen atom in each water as the center point of that molecule. At a given time, we compute the shortest distance, $r$, between a water molecule and an oxygen or carbon atom on any glycerol molecules in a mixture. We then count the number of water molecules in glycerol hydration shells from $r - \delta r$ to $r + \delta r$ and normalize this number by the number of glycerol molecules, $N_{\text{glyc}}$, in the system. The result of this analysis is denoted as the normalized hydration number, $N(r)$, as a function of distance, $r$, at various glycerol concentrations (Figure 7). Two main zones clearly show in these functions including an excluded zone and a primary hydration layer around a glycerol molecule. The excluded zone extends from the surface of a glycerol molecule, where $r = 0$, to $r \approx 2.3$ Å and the primary hydration layer extends to $r \approx 3.15$ Å, indicating a thickness of about 0.85 Å for the hydration layer. All water molecules in the primary hydration shell are referred to as the bound water and denoted as the hydration number, $N_{\text{hyd}}$. The ratio $f_b = N_{\text{hyd}} \times N_{\text{glyc}}/N_{\text{water}}$, with $N_{\text{water}}$ being the total number of water molecules in a solution, thus, indicates the fraction of water located in primary hydration shells of glycerol molecules. The results on $N_{\text{hyd}}$ and $f_b$ for all systems modeled here are included in Table 2. Specifically, for the mixture containing 4.15 mol % glycerol, $N_{\text{hyd}}$ is around 5.57, which matches well with the experimental value of 5.58 water molecules in the hydration layer of a glycerol molecule in the low concentration regime. When the glycerol concentration is increased, $N_{\text{hyd}}$ gradually decreases and at the glycerol concentration of 50 mol %, the fraction of water being bound to glycerol, $f_b$, gradually increases toward 1 (Figure 7, inset), meaning that, by average there is one water molecule per one glycerol molecule in this concentrated mixture. Furthermore, Figure 7 indicates that when the glycerol concentration is increased beyond 19 mol %, the second peak in the normalized hydration number curve disappears. This behavior is consistent with the physical picture that at high concentration, the hydration shells of adjacent glycerol molecules start to overlap.

When the hydration shells start to overlap, a much longer relaxation time emerges in our simulations. As shown in Figure 6b, the average relaxation time dramatically increases and shows a clear deviation from a linear dependence on the glycerol centration when it is beyond about 10 mol %. In highly concentrated mixtures, water molecules are strongly bound to glycerol molecules, thus, the average relaxation time increase significantly. Specifically, when the glycerol concentration is increased to 50 mol % ($f_b = 0.97$), 97% of water molecules are confined in primary hydration shells (2.3 Å $< r <$ 3.15 Å) of glycerol molecules. The hydration shells of glycerol molecules strongly overlap, and all water molecules are confined in a glycerol network. The average relaxation time of water in the mixture with 50 mol % of glycerol is about 58 ps. A similar relaxation time, 50 ps, can be extracted from the relaxation autocorrelation function for the mixture containing 35 mol % of glycerol (Figure 6b on the log-linear scale). Therefore, the relaxation time of confined water is somewhere between 50 ps to 60 ps from MD simulations. This value is lower than the value of 85 ps identified experimentally through dielectric spectroscopy for water confined in a glycerol network. Again, the difference is largely due to the fact that the SPC/E model generally underestimates the relaxation times of water.

## 4. CONCLUSIONS

We have performed the dielectric spectroscopy of glycerol-water mixtures in a wide frequency range from megahertz-to-terahertz region to systematically inspect the transition from pure water towards pure glycerol. An analysis of the dielectric response of glycerol-water mixtures has revealed four distinct relaxation processes including the rotational motion of glycerol molecules with a reorientational time of ~910 ps, water confined in a glycerol network with a relaxation time of ~85 ps, water bound in the hydration layer of a glycerol molecule with a relaxation time of ~35 ps, and bulk water with a relaxation time of ~8 ps. A critical glycerol concentration of ~7.5 mol % has been identified. Below this threshold, the dielectric response of the mixture is controlled by bulk water, bound water, and glycerol. Beyond the critical concentration, confined water emerges and contributes to the dielectric response of the mixture as well. In the regime of low glycerol concentration, by average the hydration shell of a glycerol molecule consists of about 5.58 water molecules. In mixtures with higher glycerol concentration, the hydration shells start to merge and overlap, the dielectric response from bound water shows a saturation behavior whereas the dielectric response from confined water increases with an increasing glycerol concentration. The physical picture revealed from the dielectric spectroscopy is further confirmed with molecular dynamics simulations. The results provide an in-depth understanding the dynamics of water and glycerol in their mixtures and insights into the reactivity of glycerol as a common co-solvent of water.

## ACKNOWLEDGEMENTS

Authors gratefully acknowledge financial support by the Air Force Office of Scientific Research under award number FA9550-18-1-0263 and National Science Foundation (CHE-1665157). We acknowledge Advanced Research Computing at Virginia Tech for providing computational resources and technical support that have contributed to the results reported within this paper.